\documentclass[aps,prb,twocolumn,10pt,longbibliography,floatfix,nofootinbib,superscriptaddress]{revtex4-2}

\usepackage{titlesec}
\titleformat{name=\section,numberless}[block]{\normalfont\bfseries}{\thesection}{1em}{\filright}

\usepackage{physics}
\usepackage{float}
\usepackage{bbold}
\usepackage[english]{babel}
\usepackage[utf8]{inputenc}
\usepackage{amssymb}
\usepackage{amsmath}
\usepackage{dsfont}
\usepackage{multirow,tabularx,makecell}
\usepackage[pdftex]{graphicx}
\usepackage{xspace}
\usepackage{dsfont}
\usepackage{bm}
\usepackage{nicefrac}
\usepackage[export]{adjustbox}
\usepackage{makecell}

\usepackage{tabularx} 
\usepackage{makecell} 
\usepackage{graphicx} 
\usepackage{array}    

\usepackage[pdfencoding=auto, psdextra]{hyperref}
\hypersetup{
    colorlinks,%
    citecolor=blue,%
    filecolor=blue,%
    linkcolor=blue,%
    urlcolor=blue
}

\usepackage[usenames, dvipsnames]{xcolor}

\newcommand{\xin}{x_{\mathrm{in}}}

\newcommand{\xsat}{x_{\mathrm{sat}}}

\newcommand{\transferfunc}{\mathcal{H}}
\newcommand{\ratio}{\Gamma}

\newcommand{\wc}{\omega_{\mathrm{c}}} 
\newcommand{\ws}{\omega_{\mathrm{s}}} 

\newcommand{\logten}{\log_{\mathrm{10}}} 

\begin{document}

\title{RC circuit based on magnetic skyrmions}

\author{Ismael Ribeiro de Assis}
\affiliation{Institut f\"ur Physik, Martin-Luther-Universit\"at Halle-Wittenberg, D-06099 Halle (Saale), Germany}

\author{Ingrid Mertig}
\affiliation{Institut f\"ur Physik, Martin-Luther-Universit\"at Halle-Wittenberg, D-06099 Halle (Saale), Germany}

\author{B{\"o}rge G{\"o}bel}

\affiliation{Institut f\"ur Physik, Martin-Luther-Universit\"at Halle-Wittenberg, D-06099 Halle (Saale), Germany}

\date{\today}

\begin{abstract}
Skyrmions are nano-sized magnetic whirls attractive for spintronic applications due to their innate stability. They can emulate the characteristic behavior of various spintronic and electronic devices such as spin-torque nano-oscillators, artificial neurons and synapses, logic devices, diodes, and ratchets. Here, we show that skyrmions can emulate the physics of an RC circuit—the fundamental electric circuit composed of a resistor and a capacitor—on the nanosecond time scale. The equation of motion of a current-driven skyrmion in a quadratic energy landscape is mathematically equivalent to the differential equation characterizing an RC circuit: the applied current resembles the applied input voltage, and the skyrmion position resembles the output voltage at the capacitor. These predictions are confirmed via micromagnetic simulations. We show that such a skyrmion system reproduces the characteristic exponential voltage decay upon charging and discharging the capacitor under constant input. Furthermore, it mimics the low-pass filter behavior of RC circuits by filtering high-frequencies in periodic input signals. Since RC circuits are mathematically equivalent to the Leaky-Integrate-Fire (LIF) model widely used to describe biological neurons, our device concept can also be regarded as a perfect artificial LIF neuron.
\end{abstract}

\maketitle


\section{Introduction}

Magnetic skyrmions~\cite{nagaosa2013topological, gobel2021beyond,bogdanov1989} are a prominent focus of contemporary research. They are whirl-like spin textures stabilized by the interplay of interactions, such as the Dzyaloshinskii-Moriya interaction (DMI), the exchange interaction,  and the perpendicular magnetic anisotropy (PMA). Their non-collinear magnetic moments give rise to a topological protection, quantified by the topological charge, making it possible for these textures to be observed even at room temperature~\cite{muhlbauer2009skyrmion,yu2010real,saha2024high}. 

Skyrmions are interesting fundamentally and for applications due to their intrinsic stability and efficient dynamics. Once stabilized in ferromagnetic materials (FM), skyrmions can be manipulated using current-induced torques such as the spin-transfer torque (STT)~\cite{zhang2004roles,sampaio2013nucleation}. The low current densities required to move them~\cite{jiang2017direct} and their particle-like behavior have led to the proposal of using skyrmions as information carriers in devices like the racetrack memory~\cite{parkin2004shiftable,fert2013skyrmions}, skyrmion-based applications for unconventional computing~\cite{grollier2020neuromorphic,khodzhaev2023analysis,li2021magnetic,chen2018compact,li2021magnetic}, reservoir computing~\cite{everschor2024topological,msiska2023audio, pinna2020reservoir,raab2022brownian}, neuromorphic computing ~\cite{song2020skyrmion,de2023biskyrmion,liang2020spiking}, ratchet devices~\cite{gobel2021skyrmion,souza2021skyrmion}, spin-torque oscillators~\cite{de2024circular,shen2019spin,garcia2016skyrmion} and others. This makes skyrmions promising candidates for energy-efficient alternatives to currently used technologies.

While magnetic skyrmions have been successfully employed to emulate various electronic components, such as diodes~\cite{jung2021magnetic,bindal2023antiferromagnetic,zhao2020ferromagnetic}, logic gates~\cite{zhang2015magnetic,luo2018reconfigurable,liang2021antiferromagnetic}, and transistors~\cite{zhang2015magnetictransistor, yang2023magnetic,zhao2018single}, they have not yet been directly associated with one of the most fundamental electronic components: the RC circuit, consisting of a resistor and a capacitor. The latter can be charged and discharged, and it acts as a low-pass filter when alternating currents are applied. The literature about emulating this electronic circuit with magnetic textures is scarce. Ref.~\cite{chen2015rc} considers domain walls, whose position mimics an RC circuit under the influence of external and demagnetizing fields. Yet, the potential of using skyrmions in applications mimicking this traditional electronic component has remained unexplored.

\begin{figure*}[]
  \centering
    \includegraphics[width=\textwidth]{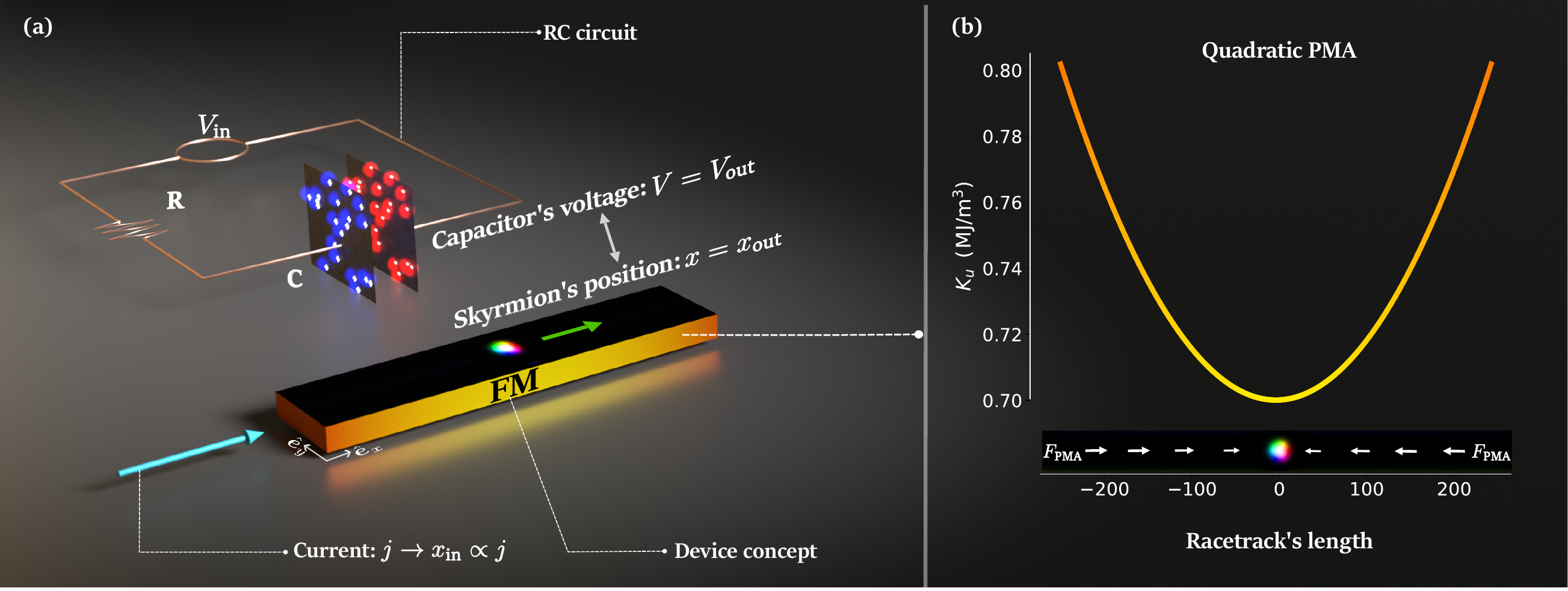}
    \caption{Schematics of the device concept. (a) Top: Artist view of an RC circuit consisting of a resistor with resistance $R$ and a capacitor with capacitance $C$. The voltage at the capacitor follows the behavior discussed in Sec.~\ref{sec:RC_circuits}. Bottom: Schematic representation of the device concept introduced in Sec.~\ref{sec:skyrmions_as_RC} featuring a Néel skyrmion stabilized at the center of a racetrack. The racetrack is composed of a ferromagnetic (FM) layer with an engineered spatial variation of the perpendicular magnetic anisotropy (PMA), $K_{\mathrm{u}}(x)$. The skyrmion can be moved along the length of the racetrack (indicated by the green arrow) via spin-transfer torques (STT) induced by an applied current (indicated by the blue arrow). The position of the skyrmion is analogous to the voltage across a capacitor in the corresponding RC circuit. The colors in the FM layer represent variations in the PMA. (b) This panel illustrates the variation of $K_{\mathrm{u}}$ across the FM layer, as shown in panel (a). The anisotropy $K_{\mathrm{u}}$ varies quadratically. The skyrmion can move in both directions by changing the sign of the applied current, with the linear force $F_\mathrm{PMA}$ (white arrows) resulting from the PMA variation pointing towards the center of the track. }
    \label{fig:device_concept}
\end{figure*}

In this paper, we establish that skyrmions can be associated with classical RC circuits, thereby opening a new avenue for skyrmion-based applications. We show analytically that the time-evolution of the position $x$ of a skyrmion driven by currents in an engineered ferromagnetic layer is equivalent to the time-dependent voltage $V$ across the capacitor in an RC circuit, as schematically summarized in Fig.~\ref{fig:device_concept}(a). Furthermore, we demonstrate that a quadratic spatial dependence of the magnetic anisotropy parameter enables us to directly map the skyrmion's equation of motion onto the differential equation for the voltage of the capacitor. Through micromagnetic simulations, we confirm that the device concept displays the typical charging and discharging dynamics with a time response in the nanosecond range. Additionally, it exhibits the signal filtering features of RC circuits functioning as low-pass filters at GHz frequencies. To our knowledge, this is the first demonstration of a direct mapping between a skyrmion system and a classical RC circuit. 

This paper is organized as follows. In Sec.~\ref{sec:RC_circuits}, we review the fundamental concepts of RC circuits. Subsequently, Sec.~\ref{sec:skyrmions_as_RC} presents our device concept. In Sec.~\ref{sec:analytical}, we present and discuss the analytical results, where we show that the differential equations characterizing the skyrmion's position $x$ and the capacitor's voltage $V$ are analogous. This step establishes the analogy $V\leftrightarrow x$ used throughout this paper. We verify these results through micromagnetic simulations in Sec.~\ref{sec:micromagnetic_simulations} and
demonstrate the typical features of a low-pass filter and show its application to transform waveforms depending on the frequency in Sec.~\ref{sec:micromagneticExample}. We conclude in Sec.~\ref{sec:conclusion}.

\section{RC circuits in electronics}
\label{sec:RC_circuits}

In this section, we briefly review the key aspects of the well-known RC circuit and introduce dimensionless equations that allow us to compare it directly with a skyrmion system later in this paper.

\begin{figure}[]
  \centering
    \includegraphics[width=\columnwidth]{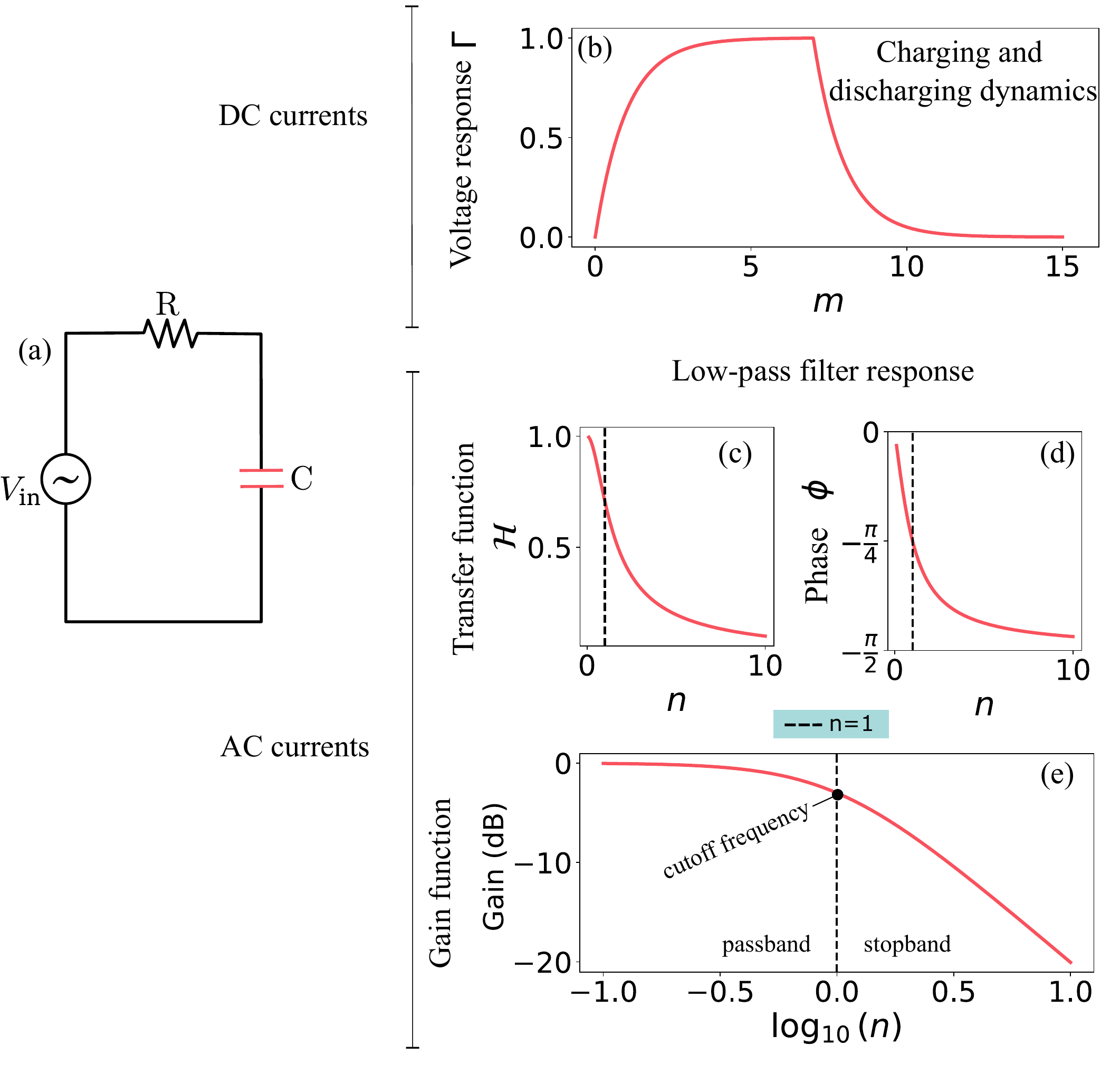}
    \caption{Characterization of RC circuits. (a) Schematic of an RC circuit with resistance $R$ and capacitance $C$ in series. (b) The charging and discharging processes under DC voltage are depicted by the normalized voltage response $\ratio = V_{\mathrm{out}}(t)/V_{\mathrm{in}}(t)$, plotted as a function of $m = t/\tau$, where $t$ is the time, and $\tau$ is the circuit's time constant. The input voltage is turned on at $m = 0$, starting the charging process of the capacitor. At $m = 7$, the current is turned off, and the capacitor discharges. (c) RC circuits are low-pass filters, attenuating high-frequency components from input signals such as AC currents. The normalized transfer function $\mathcal{H} = V_{\mathrm{out}}^A / V_{\mathrm{in}}^A$ [Eq.~\eqref{eq:ratio_RC}], is plotted versus $n = \omega/\omega_c$, where $\omega$ is the input signal frequency and $\omega_c$ is the cutoff frequency. (d) The phase shift $\phi$ [Eq.~\eqref{eq:phase_RC}] of the output signal in response to AC currents is plotted versus $n$. In (e), the gain function [Eq.~\eqref{eq:gain_RC}] is plotted against $\logten{n}$. The black dashed line in panels (c-e) marks the cutoff frequency, where $n=1$.}
    \label{fig:RC_circuit}
\end{figure}

An RC circuit consists of a capacitor with capacitance $C$ and a resistor with resistance $R$ configured in series, connected to a time-dependent `input' voltage source $V_\mathrm{in}(t)$ as depicted in Fig.~\ref{fig:RC_circuit}(a). From Kirchoff's rules, the `output' voltage $V_\mathrm{out}(t)$ across the capacitor is determined as
\begin{equation}
\label{eq:RC}
    \tau_\mathrm{c} \frac{\mathrm{d} V_\mathrm{out}(t)}{\mathrm{d} t} = -V_\mathrm{out}(t) + V_\mathrm{in}(t),
\end{equation}
where $\tau_\mathrm{c} = R C$ is the circuit's time constant, determining how quickly the circuit reacts to a change in the applied voltage.

\subsection{Behavior under direct currents (DC)}

For a constant input voltage $V_\mathrm{in}(t)=V_\mathrm{in}^0$ starting from an uncharged capacitor, one obtains the voltage response ratio $\ratio(t) = V_\mathrm{out}(t)/V_\mathrm{in}(t)$ as
\begin{equation}
    \label{eq:ratio_RC}
      \ratio =  1 - e^{-m}.
\end{equation}
We have normalized the exponent in units of the time constant, $m = t/ \tau_\mathrm{c}$. The voltage across the capacitor increases when the input voltage is applied due to the accumulation of charges on the capacitor's plates. As illustrated in Fig.~\ref{fig:RC_circuit}(b), the voltage increases up to a saturation voltage equal to $V_{\mathrm{in}}^0$. The voltage increase follows an exponential function characterized by the circuit's time constant $\tau_\mathrm{c}$. After a time of $t=5\tau_\mathrm{c}$, the voltage reaches $\sim 99\%$ of the saturation value. Once the input voltage is turned off (in our simulation at $t=7\tau_\mathrm{c}$), the capacitor discharges and follows a declining exponential function towards zero voltage.

\subsection{Behavior under alternating currents (AC)}

Another characteristic feature of RC circuits is their behavior under alternating currents (AC), specifically their application as low-pass filters. For a sinusoidal input voltage $ V_\mathrm{in}(t) = V_\mathrm{in}^\mathrm{A} \sin (\omega t)$, where $V_\mathrm{in}^\mathrm{A}$ is the amplitude and $\omega$ the frequency, the output voltage at the capacitor oscillates with the same frequency and with an amplitude $V_{\mathrm{out}}^\mathrm{A}$. This amplitude decreases, especially for $\omega > \wc$, where $\wc = \tau_\mathrm{c}^{-1}$ is the cutoff frequency -- the frequency at which the output voltage amplitude drops to $V_\mathrm{in}^\mathrm{A}/\sqrt{2}$. It marks the frequency beyond which the circuit significantly attenuates higher-frequency signals. 

It is common to describe this attenuation by the so-called `transfer function'~\cite{william2007engineering,horowitz1989art} 
\begin{equation}
\label{eq:transferfunc_RC}
    \transferfunc = \frac{V_{\mathrm{out}}^\mathrm{A}}{V_\mathrm{in}^\mathrm{A}} = \frac{\wc}{\sqrt{\wc^2+ \omega^2}} = \frac{1}{\sqrt{1+ n^2}},
\end{equation}
where we have introduced the normalized frequency $n = \omega/\wc$. This quantity can be derived from Eq.~\eqref{eq:RC} with the ansatz $V_{\mathrm{out}}(t) = V_\mathrm{in}^\mathrm{A} \sin{(\omega t + \phi_{\mathrm{c}})}$~\cite{william2007engineering}. The transfer function is plotted in Fig.~\ref{fig:RC_circuit}(c); it decreases significantly with the frequency. A related quantity called `gain function' is defined by $\mathrm{Gain} = 20 \logten |\transferfunc|$, or
\begin{equation}
\label{eq:gain_RC}
    \mathrm{Gain} = 20 \logten \frac{1}{\sqrt{1 + n^2}},
\end{equation}
where $\mathrm{Gain}$ has the unit of decibels (dB) and is typically plotted versus $\logten(n)$, as illustrated in Fig.~\ref{fig:RC_circuit}(e). In the frequency range below $\wc$ (passband), the plot is nearly constant, close to a value of 0 dB, indicating no signal attenuation. Above $\wc$ (stopband), the gain function decreases sharply, indicating a significant attenuation of $V_{\mathrm{out}}^\mathrm{A}$.

The phase acquired by the output voltage $V_\mathrm{out}(t)$ compared to the input voltage $V_\mathrm{in}(t)$ is another characteristic feature related to the oscillating charging and discharging of the capacitor in an RC circuit. This phase depends on the frequency as
\begin{equation}
\label{eq:phase_RC}
    \phi = -\arctan{\frac{\omega}{\wc}}= - \arctan{n}.
\end{equation}
Its magnitude increases at higher frequencies, as depicted in Fig.~\ref{fig:RC_circuit}(d).
In the limit of high frequencies $\omega\gg\wc$, $\phi \rightarrow -\frac{\pi}{2}$ which means that the output lacks behind the input by up to a quarter of the period. For very slow frequencies $\omega\ll\wc$, the voltage at the capacitor oscillates in phase with the input voltage.

In the following sections, we demonstrate that a ferromagnet hosting a skyrmion can be engineered to have the same behavior under AC and DC currents. We compare this system with the RC circuit using the quantities $\Gamma_s$, $\transferfunc_{\mathrm{s}}$, $\mathrm{Gain}_s$ and $\phi_s$ defined analogously to $\Gamma$, $\transferfunc$, $\mathrm{Gain}$ and $\phi$ introduced in this section.


\section{Device concept: RC-circuit analogue based on magnetic skyrmions}
\label{sec:skyrmions_as_RC}



Before we show the analytical and numerical results, we want to introduce the system unter consideration. Throughout this paper, we relate skyrmions to RC circuits by the following analogy: $V\leftrightarrow x$. 

Mathematically speaking, the \textit{position} $x$ will take the role of the \textit{voltage} $V$. Therefore, we can define the following analogies: (a) The `output' voltage $V_\mathrm{out}$ at the capacitor becomes the `output' position $x_\mathrm{out}$ of the skyrmion. (b) The `input' voltage $V_\mathrm{in}$ of the RC circuit becomes the `input' position $x_\mathrm{in} $, a quantity that will be defined as proportional to the driving current.

The equivalence of the two seemingly far-apart systems—an RC circuit and a skyrmion racetrack—is only given if the skyrmion moves in a quadratic energy landscape.
Fig.~\ref{fig:device_concept}(a) depicts our device concept, featuring a nanotrack made of a ferromagnetic (FM) layer. A Néel skyrmion is stabilized at the center, confined by the following engineered potential: It varies spatially via the perpendicular magnetic anisotropy (PMA) parameter $K_{\mathrm{u}}$, which has a quadratic dependence on the position along the $x$ direction; cf. Fig.~\ref{fig:device_concept}(b). At the center, $K_{\mathrm{u}}$ is at its minimum. Such a variation of the anisotropy parameter could potentially be realized experimentally via focused ion beam irradiation ~\cite{juge2021helium,kern2022deterministic,ahrens2023skyrmions} or a thickness modulation \cite{yu2016room, gowtham2016thickness}.

In the following Section~\ref{sec:analytical}, we map the skyrmion's equation of motion onto the differential equation for the voltage of the capacitor in an RC circuit [Eq.~\eqref{eq:RC}]. We derive the presented analogy of the voltage $V$ in the RC circuit and the position $x$ in the skyrmion system mathematically. We discuss how the energy landscape affects the skyrmion motion and how it has to be engineered to mimic the reversing force caused by the accumulation of charges on the capacitor plates. Following this mathematical framework, we validate these findings through micromagnetic simulations of the proposed device concept in Sec.~\ref{sec:micromagnetic_simulations} and discuss the physical behavior of the skyrmion in detail.


\section{Mathematical equivalence between skyrmion systems and RC circuits}
\label{sec:analytical}

This section introduces the effective equation of motion for the skyrmion, namely, the Thiele equation~\cite{thiele1973steady}. We then establish the analogy with the RC circuit by comparing the differential equations for the skyrmion's position and the capacitor's voltage.

\subsection{Skyrmion's equation of motion}
\label{sec:intro_thiele_equation}

The velocity $\bm{v}$ of a magnetic skyrmion can be described via the Thiele equation~\cite{thiele1973steady} given by 
\begin{equation}\label{eq:Thiele}
    b \bm{G} \times \bm{v} + b \underline{\bm{D}}  \alpha \bm{v} + \bm{F}_\mathrm{STT} + \bm{F}_{\mathrm{grad}} = 0,
\end{equation}
where $\alpha$ is the Gilbert damping and $b = M_{\mathrm{s}} d_z / \gamma_e$ is a constant defined by the saturation magnetization $M_{\mathrm{s}}$ of the ferromagnet,  its thickness $d_z$ and the gyromagnetic ratio $\gamma_e$.
 
Each term has the dimension of a force. The first term is the so-called gyroscopic force, characterized by the gyroscopic vector $\bm{G} =- 4 \pi N_\mathrm{Sk} \hat{\bm{e}}_z$ arising from the skyrmion's topological charge $N_\mathrm{Sk}= +1$. The second term is the dissipative force caused by the Gilbert damping. The dissipative tensor 
$D_{ij} =- \int ( \partial_{x_i} \bm{m} \cdot \partial_{x_j} \bm{m} ) \, \mathrm{d}^2 r$
only has non-zero $D_{xx}=D_{yy}\equiv - D_0$ elements due to the rotational symmetric shape of the skyrmion that is a good approximation for the investigated system throughout this paper. 

By injecting a charge current $\bm{j}$ along the track into the FM, the conduction electrons become partly spin-polarized and interact with the non-collinear magnetic texture of the skyrmion, leading to a spin-transfer torque that ultimately results in the force $\bm{F}_\mathrm{STT}$ accounted for by the third term. The STT-induced force has a gyroscopic- and dissipative-like contribution~\cite{thiaville2005micromagnetic,zhang2004roles}. Thus, Eq.~\eqref{eq:Thiele} can be rewritten as 
\begin{equation}
\label{eq:Thiele_STT}
    b \bm{G} \times (b_J \bm{j} + \bm{v}) + b \underline{\bm{D}} (\xi  b_J \bm{j} + \alpha \bm{v})  +  \bm{F}_{\mathrm{grad}} = 0,
\end{equation}
where $\bm{j}$ is the applied current density that acts along $x$ and $b_J$ is defined via $b_J = \mu_B P / e M_s (1+\xi^2)$, where $\mu_B$ is the Bohr-magneton, $P$ the current's spin polarization, $e$ the electron's charge and $\xi$  the non-adiabaticity parameter of the STT~\cite{zhang2004roles}.

The last term is governed by the potential landscape of the skyrmion. We distinguish two contributions to $\bm{F}_{\mathrm{grad}}$: (i) The repulsive force exerted on the skyrmion by the edges, where a potential gradient emerges that causes the force $\bm{F}_e$. As long as the skyrmion is close to the equilibrium position, this force acts along $y$ due to the elongated shape of the device. (ii) The force $\bm{F}_\mathrm{PMA}$ acting along $\pm x$ [cf.Fig.~\ref{fig:device_concept} (b)] caused by the change in the skyrmion's energy landscape due to the spatially engineered variation of the PMA. The gradient of the corresponding energy exerts a force on the skyrmion, which moves it towards the position with the lowest anisotropy, as illustrated in Fig.~\ref{fig:device_concept}. These forces $\bm{F}_{\mathrm{grad}}=F_\mathrm{PMA}\hat{\bm{e}}_x+F_e\hat{\bm{e}}_y$ will be discussed in more detail below.

Due to the topological charge of the skyrmion entering the gyroscopic force in the Eq.~\eqref{eq:Thiele}, the skyrmion experiences a force component transverse to the applied current (along $x$), which pushes the skyrmion toward one of the edges (along $y$), resulting in a non-zero $v_y$. This effect is known as the skyrmion Hall effect in the literature~\cite{jiang2017direct, litzius2017skyrmion}. However, the edge interaction introduces a force $F_e$ in the opposite direction of the transverse component and increases as the skyrmion approaches the edge. If the skyrmion's velocity is sufficiently small, there is a critical $y$ position for which the edge force cancels the transverse gyroscopic force component, leading to $v_y = 0$. At this point, the skyrmion moves effectively one-dimensionally along $x$. The edge confinement has already been used experimentally to guide the motion of skyrmions in devices \cite{raab2022brownian,mallick2024driving,du2015edge}. In the following calculations, we use this confinement and assume a fully one-dimensional motion along the positive $x$ direction. This is an approximation because the skyrmion is not initially at the critical $y$ coordinate. However, as our micromagnetic simulations in Sec.~\ref{sec:micromagnetic_simulations} will show, it is valid for all scenarios presented in this paper. This is why we have introduced the \textit{scalar} position $x$ as the analogous quantity to the voltage at the capacitor.

Therefore, the Thiele equation can easily be solved for the skyrmion velocity
\begin{equation}
    \label{eq:vx_edges}
    \frac{\mathrm{d} x(t)}{\mathrm{d} t} = \frac{F_{\mathrm{PMA}}(x(t))}{b D_0 \alpha} + \frac{\xi}{\alpha}b_J j.
\end{equation}
where we have isolated $\frac{\mathrm{d} x(t)}{\mathrm{d} t}$ by setting $v_y = 0$.

Lastly, we have to insert the $x$ dependence of the PMA-induced force defined as $F_{\mathrm{PMA}}(x) = -\partial_x E_{\mathrm{total}}(x)\approx -\partial_x E_{\mathrm{PMA}}(x)$, where $E_{\mathrm{total}}$ is the sum of all the interactions present in the system and $E_{\mathrm{PMA}}(x) \approx K_\mathrm{u}(x)\cdot \int [1 -  m_{z}(\bm{r})^2]\mathrm{d}^3r$
is the PMA energy. Due to the quadratic dependence of the anisotropy parameter $K_{\mathrm{u}}(x)$, the dominant contribution to $F_{\mathrm{PMA}}$ comes from the anisotropy energy $E_{\mathrm{PMA}}$ that is quadratic on $x$. Therefore, we can write $E_{\mathrm{total}} \approx A_2 x^2$, where $A_2$ is a coefficient. This assumption is valid as long as the skyrmion does not deform significantly \cite{de2023skyrmion}. 

To determine the coefficient $A_2$,  we can fit $E_{\mathrm{total}}$ versus the skyrmion position. Note that the fitted coefficient $A_2^\mathrm{fit}$ is determined by the skyrmion profile, which, in turn, is determined by the magnetic parameters and the edge interaction. The coefficient can also be determined as $A_2^\mathrm{approx}=\frac{\pi d_z r_0^2}{2}\cdot K_2$ where $K_2$ determines the spatial variation of the PMA as $K_\mathrm{u}(x)=K_\mathrm{u,min}+K_2\cdot x^2$. Due to the edges, there exists a small deviation of $A_2^\mathrm{approx}$ from $A_2^\mathrm{fit}$.  If we use $A_2^\mathrm{approx}$, Eq.~\eqref{eq:vx_edges} becomes 
\begin{equation}
\label{eq:dxdt_edges}
    \frac{\mathrm{d} x(t)}{\mathrm{d} t} = -\frac{\pi d_z r_0^2 K_2}{b D_0 \alpha}x(t) + \frac{\xi}{\alpha}b_J j(t).
\end{equation}

Equation ~\eqref{eq:dxdt_edges} provides a simple description of the skyrmion dynamics along the edges under the influence of the quadratic energy landscape and an applied current. We now compare this equation with the differential equation for the capacitor's voltage in an RC circuit and show their equivalence.

\subsection{Mapping the skyrmion device to the RC circuit}
\label{sec:map_thiele_eq}

Mapping the Thiele equation onto the differential equation characterizing an RC circuit provides a framework for comparing the skyrmion dynamics to the well-understood RC circuit. By rearranging the terms in Eq.~\eqref{eq:dxdt_edges}, we obtain
\begin{equation}
   \tau_{\mathrm{s}} \frac{\mathrm{d} x(t)}{\mathrm{d} t} = -x(t) + r_{\mathrm{s}} j(t)
\end{equation}
in terms of physical quantities: current density $j$ and skyrmion position $x$. Mathematically, we can express the equation in terms of `input' position $x_\mathrm{in}=r_\mathrm{s}j$ and `output' position $x_\mathrm{out}=x$ 
\begin{equation}
\label{eq:Thiele_RC}
   \tau_{\mathrm{s}} \frac{\mathrm{d} x_\mathrm{out}(t)}{\mathrm{d} t} = -x_\mathrm{out}(t) + x_\mathrm{in}(t).
\end{equation}
The time constant of the skyrmion system is 
\begin{equation}
\label{eq:tau}
   \tau_{\mathrm{s}} = \frac{b D_0 \alpha}{\pi d_z r_0^2 K_2} 
\end{equation}
and $r_\mathrm{s}=\tau_\mathrm{s}\frac{\xi}{\alpha}b_J$. The subscript $\mathrm{s}$ is added to refer to the skyrmion system.

This first-order differential equation~\eqref{eq:Thiele_RC} for the skyrmion position $x$ is equivalent to the differential equation for the voltage $V$ of a capacitor in an RC circuit [Eq.~\eqref{eq:RC} in Sec.~\ref{sec:RC_circuits}]. This is the main result of this paper, as both equations must have the same solutions. We will discuss them in detail in the rest of the paper. Here, it becomes clear that a direct mapping was only possible due to the quadratic $x$ dependence of the PMA, which leads to the linear term $-x_{\mathrm{out}}(t)$ on the right-hand side of the equation.

\subsection{Solution for direct currents (DC)}

Equation~\eqref{eq:Thiele_RC} has the same solution as Equation~\eqref{eq:RC}. Therefore, to describe the skyrmion's behavior under direct currents $j(t)=j^0$, i.\,e., a constant input $x_\mathrm{in}(t)=x_\mathrm{in}^0$, we can introduce the equivalent response ratio $\ratio_{\mathrm{s}} = x_\mathrm{out}(t)/ x_\mathrm{in}(t)$ that reads
\begin{equation}
\label{eq:ratio_skyrmion}
   \ratio_\mathrm{s} = 1 - e^{-m},
\end{equation}
where $m =t/\tau_{\mathrm{s}}$ is the normalized time defined in analogy to the RC circuit ($m =t/\tau_{\mathrm{c}}$). 

This function has already been shown in Fig.~\ref{fig:RC_circuit}(b). For comparison, the trajectory of a skyrmion based on micromagnetic simulations will be shown in Fig.~\ref{fig:skyrmion_RC} and discussed in Sec.~\ref{sec:micromagnetic_simulations}. Still, we want to briefly describe what is expected based on the above solution.

The coordinate behaves like the voltage of a charging capacitor: The skyrmion starts in the middle of the device at $x=0$. For a constant current, the skyrmion moves along $x$ following an exponential function in time. The velocity decreases steadily, and the skyrmion asymptotically approaches a critical coordinate that is equal to $x_\mathrm{in}^0 = r_\mathrm{s}j^0$ at which the driving force and the PMA-induced force compensate each other. This behavior is analogous to the charging of a capacitor. Once the current is turned off, the skyrmion returns to the initial position. This process is analogous to the discharging of a capacitor following the PMA gradient.

\subsection{Solution for alternating currents (AC)}

When alternating current is applied, the skyrmion oscillates as well. Under AC currents, the amplitude of the oscillating voltage at the capacitor in an RC circuit is significantly attenuated ($V_\mathrm{out}^\mathrm{A}\ll V_\mathrm{in}^\mathrm{A}$) when the input frequency is above the cutoff frequency $\omega\gg\wc$ [cf. Sec.~\ref{sec:RC_circuits}]. Therefore, the amplitude of the spatial oscillation of the skyrmion also is attenuated ($x_\mathrm{out}^\mathrm{A}\ll x_\mathrm{in}^\mathrm{A}$) when the input frequency is above the skyrmion's cutoff frequency $\omega\gg\omega_\mathrm{s}$.

Physically speaking, we consider a sinusoidal current density $j(t) = j^\mathrm{A} \sin{(\omega t)}$, i.\,e., a sinusoidal `input' $x_\mathrm{in}(t) = x_\mathrm{in}^\mathrm{A} \sin{(\omega t)}$, where $\omega$ is the frequency and $j^\mathrm{A}$ is the current amplitude. The skyrmion position $x(t)$ oscillates around its initial position with an amplitude $x_{\mathrm{out}}^\mathrm{A}$ and frequency $\omega$. Moreover, a phase shift $\phi_s$ occurs between input and output.

To better describe this behavior, we utilize the quantities characterizing the low-pass filter behavior as in Sec.~\ref{sec:RC_circuits}: the transfer function $\transferfunc_{\mathrm{s}}$, the gain function $\mathrm{Gain}_{\mathrm{s}}$ and the phase shift $\phi_{\mathrm{s}}$. The frequency dependence of all three quantities has already been shown in Fig.~\ref{fig:RC_circuit}(c-e). Therefore, the functions will be compared with micromagnetic simulations in Fig.~\ref{fig:ratio_gain_skyrmion} in Sec.~\ref{sec:micromagnetic_simulations}. Below, we discuss what is to be expected based on the established mathematical equivalence to the RC circuit:

As in Section~\ref{sec:RC_circuits}, we use the ansatz $x_{\mathrm{out}}(t) = x_{\mathrm{out}}^\mathrm{A} \sin{(\omega t + \phi_{\mathrm{s}})}$, resulting in the skyrmion transfer function:
\begin{equation}
\label{eq:transferfunc_skyrmion}
    \transferfunc_{\mathrm{s}}= \frac{x_\mathrm{out}^\mathrm{A}}{x_\mathrm{in}^\mathrm{A}} = \frac{x_{\mathrm{out}}^\mathrm{A}}{r_s j^\mathrm{A}} =  \frac{\ws}{\sqrt{\ws^2+\omega^2}} = \frac{1}{\sqrt{1+n^2}},
\end{equation}
where $\ws = 1/ \tau_s$ is the skyrmion's cutoff frequency and $n=\omega/\ws$ is the normalized frequency defined in analogy to the RC circuit ($n=\omega/\omega_\mathrm{c}$). In the limit $\omega\gg\ws$, the transfer function becomes $\transferfunc_{\mathrm{s}}=1/\omega$.

From the transfer function, we define the gain function $\mathrm{Gain}_{\mathrm{s}} = 20 \logten| \transferfunc_{\mathrm{s}} |$ for the skyrmion system as
\begin{equation}
\label{eq:gain_skyrmion}
\mathrm{Gain}_{\mathrm{s}} = 20\logten \frac{\ws}{\sqrt{\ws^2+\omega^2}} = 20\logten \frac{1}{\sqrt{1+n^2}},
\end{equation}
where in the limit $\omega\gg\omega_\mathrm{s}$, it becomes $\mathrm{Gain}_{\mathrm{s}}=-20 \logten( \omega)$.

Finally, the phase shift is obtained as
\begin{equation}
\label{eq:phase_skyrmion}
    \phi_{\mathrm{s}} = -\arctan{\frac{\omega}{\omega_{s}}}.
\end{equation}
In the limit $\omega\gg\omega_\mathrm{s}$, the phase shift becomes $\phi_{\mathrm{s}} = -\pi/2$, meaning that the skyrmion position lacks behind the applied current by up to a quarter of a period.

In summary, we have established a direct connection of a skyrmion system to an electronic RC circuit. By engineering the PMA coefficient $K_{\mathrm{u}}(x)$ we have mapped the differential equation for skyrmion's position $x$ onto the equation for the capacitor's voltage $V$. As expected, the analogous quantities $x$ and $V$ show the same behavior uncer currents because both their corresponding input quantities $x_\mathrm{in}$ and $V_\mathrm{in}$ are proportional to an applied current. Next, we verify the validity of these analytical results via micromagnetic simulations.


\section{Demonstrating RC circuit behavior of skyrmions via micromagnetic simulations}

\label{sec:micromagnetic_simulations}

In this section, we present the micromagnetic simulation results, confirming the skyrmion device's mathematical analogy to the RC circuit established in the previous section. After introducing the method in Sec.~\ref{sec:method}, we demonstrate the behavior under DC currents, analogous to a capacitor's charging and discharging processes, in Sec.~\ref{sec:micromagneticDC}. Afterward, in Sec.~\ref{sec:micromagneticAC}, we analyze the behavior under AC currents and verify the device's low-pass filter behavior.

\subsection{Method: Micromagnetic simulations}\label{sec:method}

To simulate the device concept, we have used the GPU-accelerated software package MuMax3~\cite{vansteenkiste2011mumax,vansteenkiste2014design}, which solves the Landau-Lifshitz-Gilbert (LLG) equation
\begin{align}
\nonumber
\partial_t \bm{m}_i = &-\gamma_e \bm{m}_i \times \mathbf{B}^{i}_\mathrm{eff} + \alpha  \bm{m}_i \times  \partial_t \bm{m}_i
\\\nonumber
&-\frac{b_J}{M_s}[\bm{m}_i \times (\bm{m}_i \times (\mathbf{j} \cdot \mathbf{\nabla}) \bm{m}_i)]
\\
\label{eq:LLG_mumax3}
&-\xi b_J[\bm{m}_i \times (\mathbf{j} \cdot \mathbf{\nabla}) \bm{m}_i],
\end{align}
that describes the time evolution of each magnetic moment $\bm{m}_i$ which precesses around the effective field $\mathbf{B}^{i}_\mathrm{eff} = \partial E_{\mathrm{total}}/M_s \partial \bm{m}_i$. The field is derived from the system's total energy density $E_{\mathrm{total}}$, given as the sum of the interactions present in our considered system, namely, the exchange interaction $E_\mathrm{ex}=A_\mathrm{ex}\int |\nabla \bm{m}(\bm{r})|^2 \mathrm{d}^3r$, the interfacial-type Dzyaloshinskii–Moriya interaction (DMI), defined by $E_\mathrm{DMI}^{\mathrm{i}}=D_\mathrm{i}\int  [m_z(\bm{r}) \nabla \cdot \bm{m}(\bm{r}) - (\bm{m}(\bm{r}) \cdot \nabla) m_z(\bm{r})]\mathrm{d}^3r$, the perpendicular magnetocrystalline anisotropy (PMA) $E_\mathrm{PMA} = \int  K_\mathrm{u}(\bm{r}) [1 -  m_{z}(\bm{r})^2]\mathrm{d}^3r$ and the energy of demagnetization fields. The second term in Eq.~\eqref{eq:LLG_mumax3} is the Gilbert damping, with strength $\alpha$. The last two terms account for the current-induced spin-transfer torque.

We simulate an interface of Co/Pt with magnetic parameters similar to Ref.~\cite{sampaio2013nucleation}, allowing a Néel skyrmion to be stable at the center of the track. In all simulations, the FM layer has dimensions of $504$ nm × $42$ nm × $1$ nm and a cell size of $1$ nm × $1$ nm × $1$ nm. The parameter values are $M_s = 0.58$ MA/m, $\alpha =0.3$, $\xi= 0.2$, $P=0.35$. The interaction parameters are the interfacial DMI $D_\mathrm{i}= -3.5 $ mJ/m$^2$ and exchange constant $A_\mathrm{ex}= 15$ pJ/m. 

To realize the variation in the PMA parameter $K_{\mathrm{u}}(x)$, we divide the FM layer into 252 regions. The value of $K_{\mathrm{u}}$ changes at every two cells along the $x$ direction. This is because MuMax3 restricts the number of regions with different parameters to 256. At the center of the device, $K_{\mathrm{u}}$ is at its minimum, $K_{u,\mathrm{min}}$, and it increases quadratically towards $K_{u, \mathrm{max}}$ at both ends of the track, forming a parabolic profile as shown in Fig.~\ref{fig:device_concept}(b). We express this anisotropy as $K_{\mathrm{u}}(x) = K_{u,\mathrm{min}} + K_2 x^2$. The minimum value is $K_{u,\mathrm{min}} = 0.7 $ MJ/m$^3$ and the maximum value $K_{u,\mathrm{max}} = 0.8$ MJ/m$^3$ so that the quadratic spatial dependence of $K_{\mathrm{u}}(x)$ is quantified by $K_2=4(K_{u,\mathrm{max}}-K_{u,\mathrm{min}})/l^2= 1.54$ EJ/m$^5$ where $l=504\,\mathrm{nm}$ is the length of the device.

\subsection{Skyrmion under DC currents emulating a charging capacitor}\label{sec:micromagneticDC}

\begin{figure*}[]
  \centering
    \includegraphics[width=0.75\textwidth]{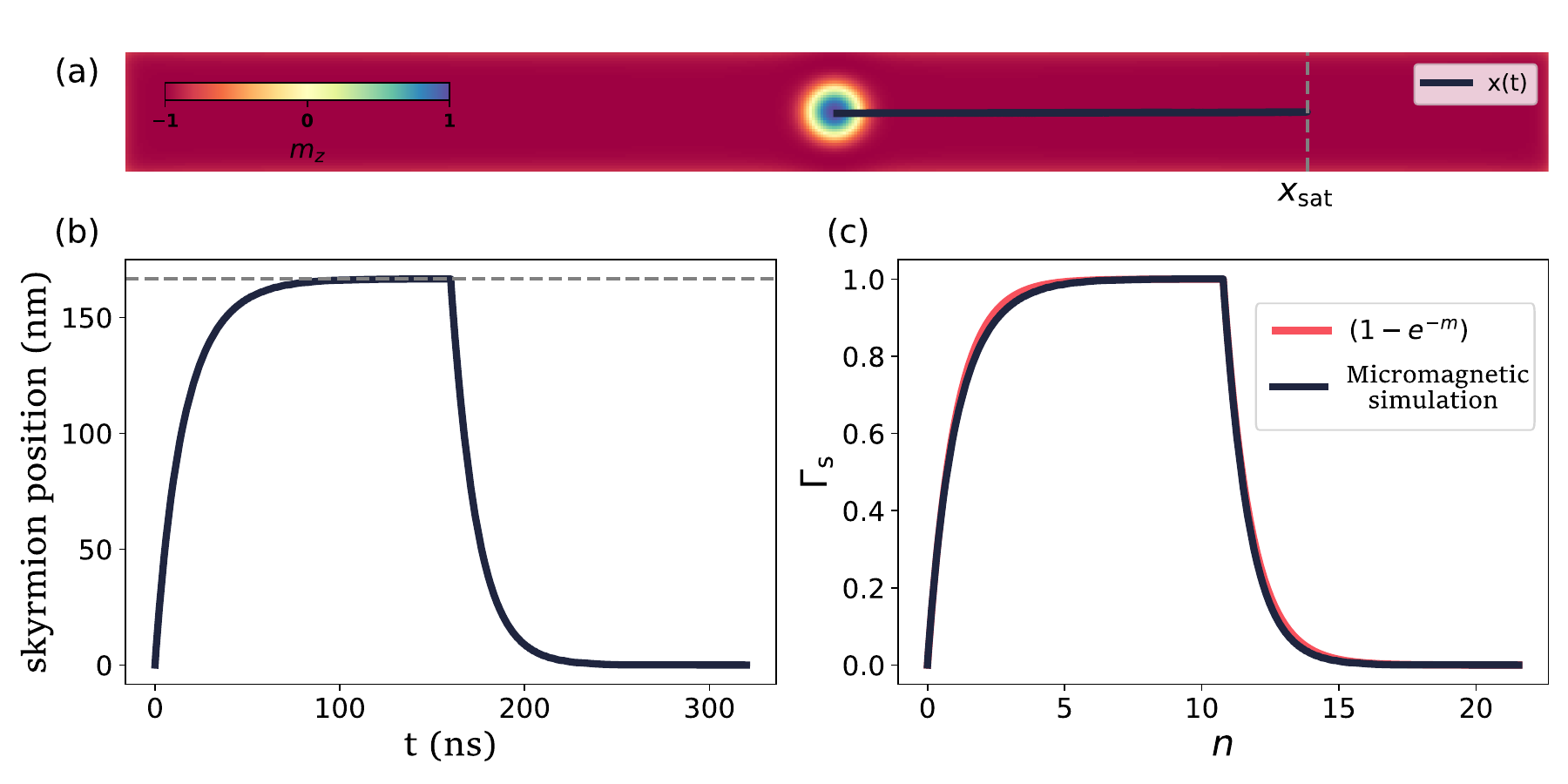}
    \caption{Comparison of micromagnetic simulations with result based on the Thiele equation. (a) Skyrmion moving in the racetrack based on micromagnetic simulations (colormap corresponds to the magnetization component $m_z$). The dark blue line indicates the skyrmion's trajectory when applying a direct current of $-0.5 \times 10^{12}$ A/m$^2$ for $t = 160$ ns and back to the origin once the current is switched off. The saturation position $\xsat = x_\mathrm{in}^0 \approx 166.5$ nm is indicated by the dashed line. (b) Time evolution of the skyrmion's position $x_\mathrm{out}(t)$ based on micromagnetic simulations, corresponding to the trajectory shown in panel (a). (c) The response $\Gamma_\mathrm{s}=x_\mathrm{out}/x_\mathrm{in}$ is plotted against $m = t/\tau$. The black curve shows the numerical data based on the micromagnetic simulation, which is proportional to the data shown in (b). The red curve resembles the exponential function derived analytically, for which the time constant $\tau = 14.84$ ns has been determined numerically, as explained in the main text. This exponential function is the same for the skyrmion system and for the RC circuit, $\Gamma_\mathrm{s}=\Gamma$, due to the established equivalence of Eq.~\eqref{eq:ratio_skyrmion} and \eqref{eq:ratio_RC}.}
    \label{fig:skyrmion_RC}
\end{figure*}

Similar to what we have discussed based on the analytical Thiele equation in Sec.~\ref{sec:analytical}, we begin our detailed analysis by showing results for constant currents. We start with a skyrmion stabilized in the center of the device, where the anisotropy is lowest. Under a constant DC current density $j(t) = j^0 = -0.5 \times 10^{12}$ A/m$^2$, the skyrmion moves along the racetrack's length, as illustrated in Fig.~\ref{fig:skyrmion_RC}(a). We apply $j^0$ for $160$ ns, then turn it off and track the skyrmion's position for an additional $160$ ns. The resulting skyrmion trajectory is plotted versus time in Fig.~\ref{fig:skyrmion_RC}(b). The confined geometry restricts the displacement along the $y$-direction, and the skyrmion moves along the edges at a constant $y$ coordinate.

The skyrmion moves effectively one-dimensionally along $x$ until a saturation position at $\xsat \approx 166.6$ nm is reached. The velocity steadily decreases, and the skyrmion asymptotically approaches the position $\xsat$. At this position, the PMA- and the STT-induced forces compensate each other. The saturation position depends linearly on the applied current. Therefore, a larger current requires a stronger PMA-induced force to counteract the increased STT-induced force. As a result, the skyrmion moves further away from the center, where the PMA-induced force is weaker. Note that without the change in the PMA parameter and the corresponding force $F_\mathrm{PMA}$, the skyrmion would keep moving with a constant velocity. Here, however, the engineered energy landscape makes the skyrmion act analogous to the charging process in an RC circuit.

Once the current is turned off, the skyrmion returns to the center of the track. Upon approaching its initial position at $x=0$, the variation of the PMA converges to zero, so the PMA-induced force, velocity, and position asymptotically approach zero. This process is analogous to the discharging process of a capacitor.

In Figure~\ref{fig:skyrmion_RC}(c), we directly compare these results of the micromagnetic simulations to the exponential time evolutions of the skyrmion position [Eq.~\eqref{eq:ratio_skyrmion}] that we have derived analytically in Sec.~\ref{sec:analytical}. Note that this equation is equivalent to the capacitor's voltage in an RC circuit [Eq.~\eqref{eq:ratio_RC}] based on the response ratio $\ratio$. As a result, the micromagnetic simulations agree well with the exponential function.

\begin{figure*}[t!]
  \centering
    \includegraphics[width=\textwidth]{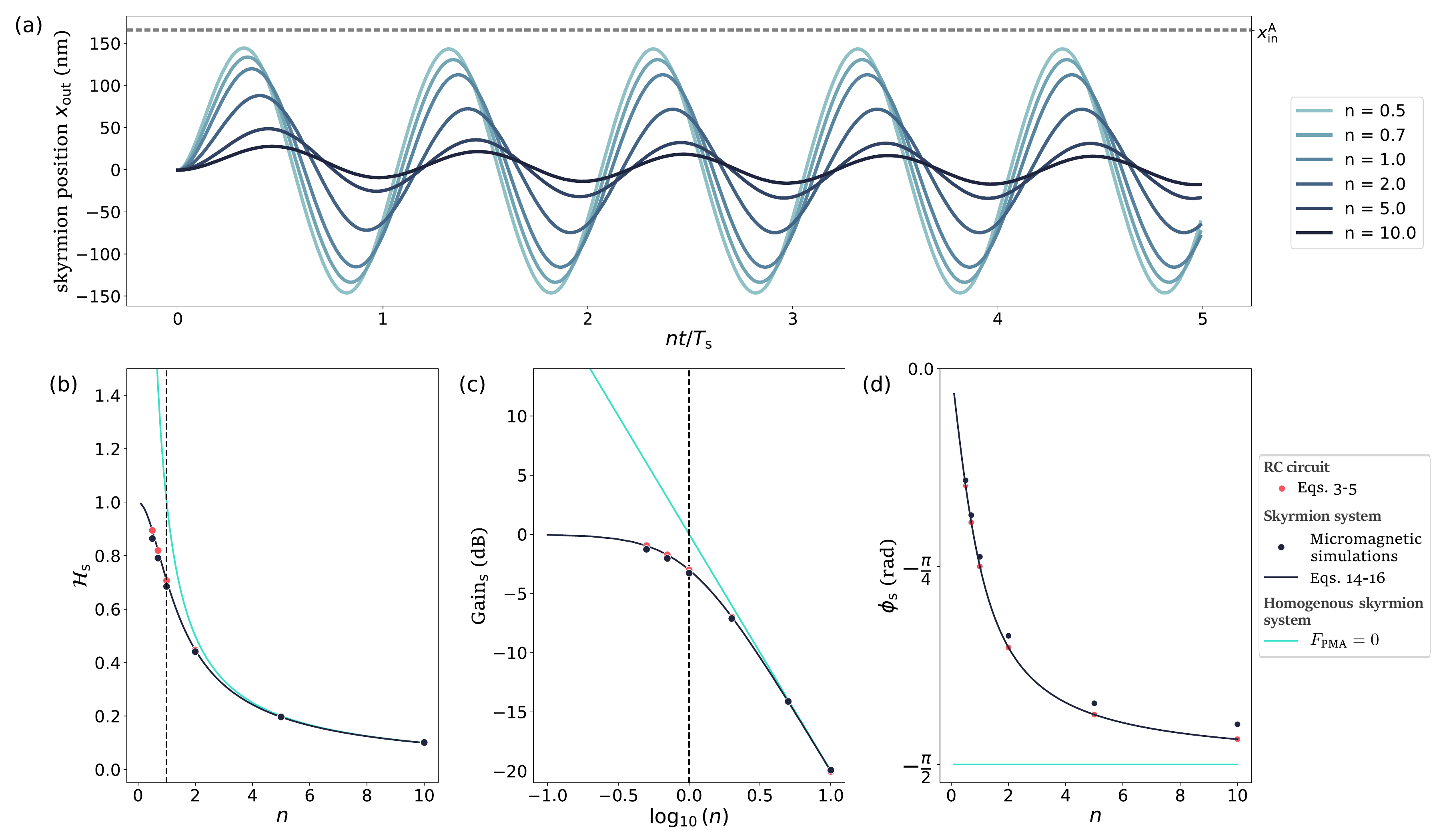}
    \caption{Alternating current (AC) behavior and low-pass filter dynamics of the skyrmion device concept.
(a) Trajectory of the skyrmion $x_\mathrm{out}(t)$ driven by currents with different frequencies (different colors, normalized as $n = \omega / \omega_s$) based on micromagnetic simulations. The current amplitude is $j^\mathrm{A}=0.5 \times 10^{12}$ A/m$^2$. The dashed line marks the saturation point $x_\mathrm{sat}=x_\mathrm{in}^A$ to which the skyrmion would move for a corresponding direct current (DC) with the same magnitude. (b-d) Comparison with the analytical results established in Sec.~\ref{sec:analytical}. Dark blue dots represent the micromagnetic simulations of the skyrmion system as simulated in (a). The red dots are added to represent the corresponding analytical values obtained from Eqs.~(3–5) for direct comparison with the RC circuit. The solid dark blue line represents the analytically derived behavior from the Thiele equation [Eqs.~14-16]. These lines are equivalent to the behavior of the RC circuit. 
(b) Normalized transfer function, $\mathcal{H}=x_\mathrm{out}^\mathrm{A}/x_\mathrm{in}^\mathrm{A}$, plotted against the frequency quantifying the attenuation of the oscillation amplitude. (c) Gain function plotted versus $\logten{(n)}$. 
(d) Phase shift between input current and observed oscillation of the skyrmion, respectively. The turquoise lines in (b-d) are analytical results representing a system without a PMA gradient, where the corresponding functions are obtained by setting $F_{\mathrm{PMA}} = 0$ in Eq.~\eqref{eq:vx_edges} [or $K_2 = 0$ in Eq.~\eqref{eq:dxdt_edges}]. }
\label{fig:ratio_gain_skyrmion}
\end{figure*}

For the comparison presented in Figure~\ref{fig:skyrmion_RC}(c), the time constant $\tau_\mathrm{s}$ has been determined using Eq.~\eqref{eq:tau} derived in Sec.~\ref{sec:analytical} based on the stabilized skyrmion in the micromagnetic simulations. We have numerically determined the coefficient $D_0$ by integrating over the stabilized magnetic texture $\bm{m}(\bm{r})$ resulting in $D_0 \approx 15.9$ and the skyrmion radius $r_0 \approx 15.52$ nm. To improve the numerical accuracy, we have not used the analytically derived approximation for the coefficient $A_2^\mathrm{approx}=\frac{\pi d_z r_0^2}{2}K_2\approx 0.6$ $\mu$J/m$^2$ but have fitted the total energy against the skyrmion position resulting in $A_2^\mathrm{fit} \approx 0.53$ $\mu$J/m$^2$. Based on these quantities, the time constant is $\tau_s \approx 14.84$ ns, characterizing the exponential function that describes the skyrmion trajectory as long as the skyrmion does not deform. Note that the device concept can operate at the nanosecond time scale. As a result, it could be an alternative to RC circuits (or analog circuits~\cite{horowitz1989art}) that face challenges in operating at this scale. Furthermore, we can determine $x_\mathrm{in}^0=r_sj^0=\tau_s\frac{\xi}{\alpha}b_Jj^0 \approx 166.5$ nm, which is very close to the numerically determined saturation position $x_\mathrm{sat}$ that the skyrmion converges to for the applied current.

The analytical curve exhibits excellent agreement with the results of the micromagnetic simulations. It was important to prove this agreement numerically because the derivation of Eq.~\eqref{eq:ratio_skyrmion} assumes a deformation-less one-dimensional motion of the skyrmion, which is an approximation that is not necessarily valid. The small deviations in Fig.~\ref{fig:skyrmion_RC}(c) are primarily due to deformations at the edge. These are captured by the dissipation tensor element $D_0$ and not accounted for by Eq.~\eqref{eq:ratio_skyrmion}. Note that if the skyrmion were to experience a more significant deformation, $D_0$ would acquire a dependency on $x$~\cite{de2023skyrmion}. This would lead to the skyrmion system emulating the behavior of a non-linear circuit, where the resistance is a function of the capacitor's voltage. Nonetheless, for the scope of this work, we emulate a classical RC circuit with very good agreement that comes from successfully suppressing such deformations, as we estimate a change of only $\sim 1.3\%$ in $D_0$ throughout the simulation. In summary, we have verified the validity of Eq.~\eqref{eq:ratio_skyrmion} via micromagnetic simulations.

\subsection{Skyrmion under AC currents acting as a low-pass filter}\label{sec:micromagneticAC}

Next, we simulate the skyrmion system under a sinusoidal current $j(t) = j^\mathrm{A}\sin(\omega t)$, where, in the context of our analogy, the input is $x_{\mathrm{in}}(t) = x_{\mathrm{in}}^\mathrm{A}\sin(\omega t)$. The skyrmion is stabilized in the center of the device, and it oscillates along the $\pm x$ direction once the AC current is applied. We examine the system under different frequencies $\omega=n\ws$. Each simulation runs for a total time of $5 T_\mathrm{s}/n$, where $T_\mathrm{s} = 2 \pi / \ws$ is the period defined by the cutoff frequency. For these simulations, we keep the parameters consistent with the values of the previous section for DC currents with $j^\mathrm{A}=j^0$. Therefore, the skyrmion's cutoff frequency $\ws =1/\tau_s \approx 67.23$ MHz.

Figure~\ref{fig:ratio_gain_skyrmion} shows how the skyrmion's oscillatory response varies with the current's frequency $\omega$.  In Fig.~\ref{fig:ratio_gain_skyrmion}(a), the skyrmion position is plotted over the time divided by the period to compare the results for different frequencies. The gray dashed line shows the saturation point $\xsat=x_\mathrm{in}^\mathrm{A}= r_\mathrm{s} j^\mathrm{A} \approx 166.5\,\mathrm{nm}$, i.\,e., the coordinate to which the skyrmion would move asymptotically if a constant current $j^0$ were applied instead. The position amplitude $x_\mathrm{out}^\mathrm{A}$ exhibits a decreasing trend with the increase in the frequency due to the variation of the PMA. We can see this behavior more clearly in the following analysis by numerically determining the transfer and gain functions based on the ratio $x_\mathrm{out}^\mathrm{A}/x_\mathrm{in}^\mathrm{A}$.

In Figures~\ref{fig:ratio_gain_skyrmion}(b-d), we analyze this low-pass filter dynamics by plotting the quantities $\transferfunc_{\mathrm{s}}$, $\mathrm{Gain}_s$ and $\phi_s$ obtained from the trajectories in Fig.~\ref{fig:ratio_gain_skyrmion}(a). Moreover, to highlight the crucial influence of the quadratic dependence of $K_{\mathrm{u}}(x)$, we compare the simulations with the analytical results of a system without a PMA gradient.

In Figure~\ref{fig:ratio_gain_skyrmion}(b) the transfer function $\transferfunc_{\mathrm{s}}=x_\mathrm{out}^\mathrm{A}/x_\mathrm{in}^\mathrm{A}$ is plotted versus $n=\omega/\omega_\mathrm{s}$ and in Figure~\ref{fig:ratio_gain_skyrmion}(c) the gain function $\mathrm{Gain}_{\mathrm{s}}=20\logten \transferfunc_{\mathrm{s}}$ is plotted versus $\logten(n)$. The dots show the results of micromagnetic simulations based on the trajectories discussed before, and the dark lines show the analytical result [Eqs.~(\ref{eq:transferfunc_skyrmion}-\ref{eq:gain_skyrmion})] derived in Sec.~\ref{sec:analytical}. For a better comparison with the RC circuit, we add the analytical values obtained from Eqs.~(\ref{eq:transferfunc_RC}-\ref{eq:gain_RC}) as the red scatter dots. The turquoise lines represent a system without a PMA gradient, discussed in detail below. The black dashed line marks the cutoff frequency where $n=1$, i.\,e., $\omega=\ws$. In Fig.~\ref{fig:ratio_gain_skyrmion}(b), for $n=1$, $x_\mathrm{out}^\mathrm{A}$ has been attenuated to $\approx 0.677 x_\mathrm{in}^\mathrm{A}$ for the micromagnetic simulations, in good agreement with the analytical value $x_\mathrm{in}^\mathrm{A}/\sqrt{2}$.  The scatter dots slightly deviate from the analytical results, primarily due to the small deformations of the skyrmion. This is because we have considered a constant $D_0 \approx 15.9$ in the Thiele equation, which is not exactly constant in all simulations.

While the transfer function provides direct insight into the ratio between the output and input amplitudes as a function of frequency, the gain function plotted in Fig.~\ref{fig:ratio_gain_skyrmion}(c) offers a clearer view of the low-pass filter behavior. The simulations closely follow the expected signal filtering dynamics, displaying a flat $\mathrm{Gain}_{\mathrm{s}}$ for low frequencies, indicating no attenuation of the output signal. At $n=1$, we have obtained $\mathrm{Gain}_{\mathrm{s}} \approx -3.28$ dB where a value of $-3.01$ dB was expected analytically. Beyond this frequency, $x_\mathrm{out}^A$ is drastically attenuated.

The significance of the PMA gradient in the skyrmion system becomes apparent when compared to a homogenous system's transfer and gain functions, i.e., a system with constant $K_{\mathrm{u}}$. We can analytically compute these quantities by setting $F_{\mathrm{PMA}} = 0$ in Eq.~\eqref{eq:dxdt_edges} yielding the transfer function $\mathrm{\mathcal{H}}_s =  \ws /\omega = 1/n$ and the Gain function $\mathrm{Gain}_s = -20 \log_{\mathrm{10}}(n)$. These are plotted in Figs.~\ref{fig:ratio_gain_skyrmion}(b-c) represented by the turquoise lines. Mathematically, these quantities behave similarly at high frequencies in both systems. The difference lies in the homogenous system's divergent behavior of $\transferfunc$ in the limit $\omega \rightarrow 0$. This suggests that a system without a PMA gradient could suppress high-frequency signals while amplifying low-frequencies. Since there is no $F_{\mathrm{PMA}}$ the skyrmion moves infinitely along the racetrack for DC currents. Under AC currents, the skyrmion oscillates uniformly in both directions depending on the current sign. 

At the high-frequency limit, the current's sign changes rapidly, leading to an amplitude attenuation. The behavior for the system with the PMA gradient is similar because at this limit ($\omega \gg \omega_\mathrm{s}$), the skyrmion's rapid displacement results in small amplitude variations, making the $K_{\mathrm{u}}(x)$ parameter's spatial dependence negligible due to its parabolic profile. As shown in Fig.~\ref{fig:ratio_gain_skyrmion}(b), the transfer function still shows a $1/\omega$ dependence, similar to the system without a PMA gradient. As a result, while both scenarios attenuate high frequencies at the same rate of $-20$ dB/decade, as shown by the gain function in Fig.~\ref{fig:ratio_gain_skyrmion}(c), the gain for a system without the PMA gradient continues to rise instead of leveling off at low frequencies. Physically speaking, at very low frequencies, even a small current would cause the skyrmion’s oscillation amplitude to exceed the length of the racetrack.

Keeping the analogy with electronic circuits, the divergent transfer function behavior of the homogenous system is analogous to that of the so-called \textit{ideal operational amplifier (Op-amp) integrators}~\cite{william2007engineering,horowitz1989art}. This theoretical circuit is known to behave as a low-pass filter for high frequencies but amplifies the output voltage at low frequencies. Such a system has a theoretically infinite gain, making it unstable and impractical to use as a stand-alone circuit; real-world circuits that approximate the ideal case (significantly high but finite gain) are typically combined with other electronic components to limit the gain or fine-tune the frequency range over which they can effectively operate~\cite{franco2002design}. Similarly, while a skyrmion system without a PMA gradient could filter high-frequency signals like a low-pass filter, it faces practical limitations at low frequencies. The unbound motion of the skyrmion would require close to an infinitely long racetrack to accommodate its divergent $x_{\mathrm{out}}^A$, which is not feasible.

Finally, another result that further indicates the low-pass filtering dynamics of our device concept is the phase difference between $x_\mathrm{out}(t)$ and $x_\mathrm{in}(t)$. The skyrmion position oscillates out of phase with the applied current, as shown in Fig.~\ref{fig:ratio_gain_skyrmion}(a). We determine the phase shift $\phi_{\mathrm{s}}$ from the simulations and plot it against $n$ in Fig.~\ref{fig:ratio_gain_skyrmion}(d) as black dots. The red dots are the theoretical values for the RC circuit [Eq.~\eqref{eq:phase_RC}]. When the variation of the PMA is large throughout the trajectory, as is the case for larger oscillation amplitudes at low frequencies, the position follows the phase of the current, and we observe almost no phase shift in the micromagnetic simulations. However, in the limit of fast frequencies, the oscillation amplitude is attenuated so much that the influence of the PMA variation is negligible, and the phase shift converges to $-\pi/2$. This is the limit of a homogenous system, where a constant phase shift is expected because of the direct proportionality of the velocity to the applied current.

In summary, we have demonstrated key features of a low-pass filter in the skyrmion device. The agreement of the micromagnetic simulations with the analytically derived result shows that the skyrmion device indeed emulates the behavior of an electronic RC circuit under direct and alternating currents. The PMA gradient is crucial because it limits the skyrmion's oscillation amplitude, allowing it to function as a true low-pass filter.

\begin{figure*}[]
  \centering    
  \includegraphics[width=\textwidth]{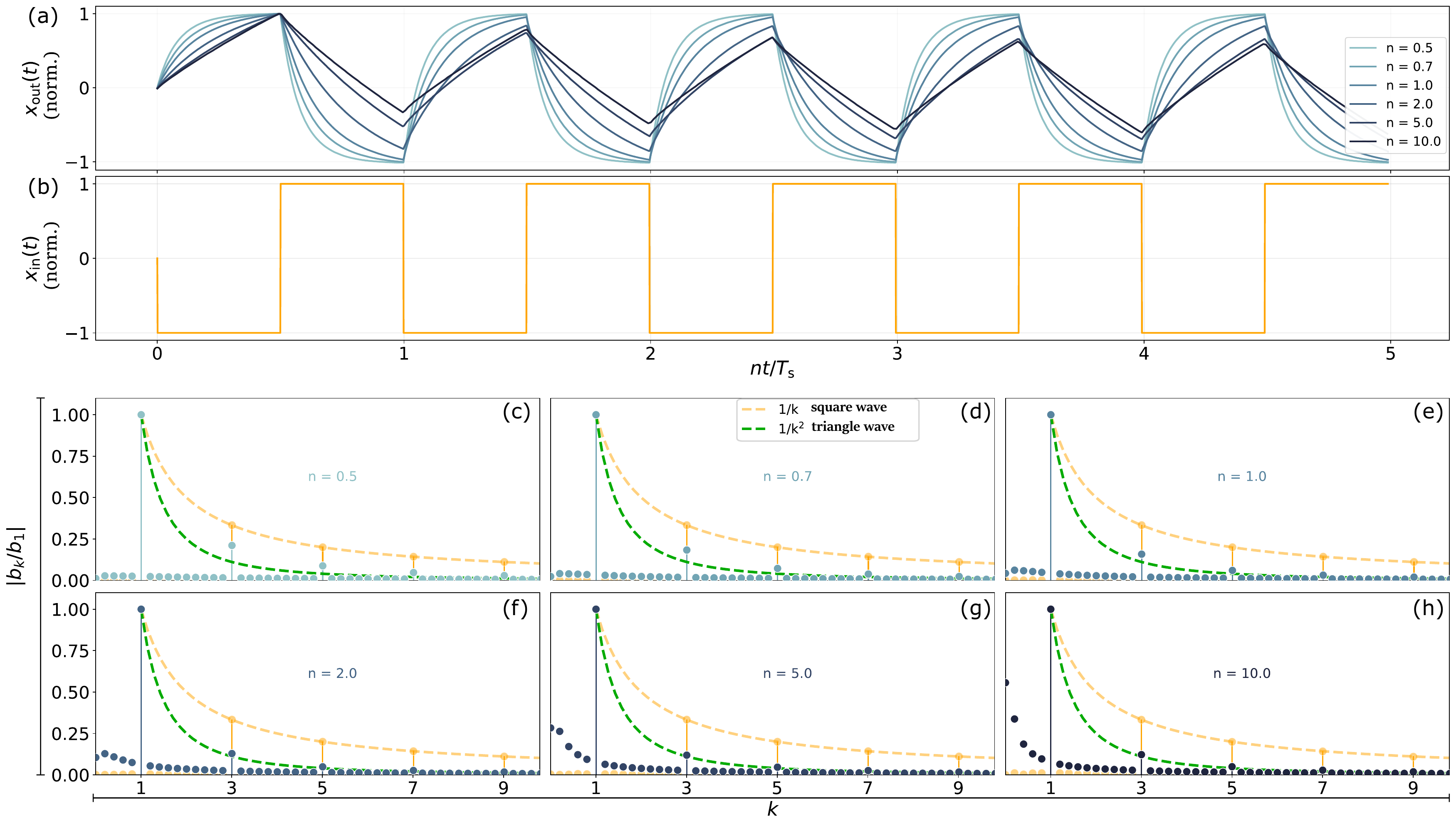}
  \caption{Response of the skyrmion device concept to square-wave current inputs.
(a) Skyrmion trajectory $x_\mathrm{out}(t)$, normalized by the oscillation amplitude $x_\mathrm{out}^\mathrm{A}$, for various driving frequencies (blue pallet colors, normalized as $n = \omega / \omega_s$). (b) Normalized input $\xin(t)$ (proportional to the square-wave current) as a function of time normalized by the period. (c-j) Fourier coefficients $b_k$ of $x_\mathrm{out}(t)$ (blue pallet scatter dots) and $\xin(t)$ (orange scatter dots) for the different current frequencies, as shown in (a) and (b). The orange and green dashed lines are added for comparison. They represent the coefficients' peak decay rates $\propto1/k$ and $\propto1/k^2$, corresponding to a perfect square and triangular wave, respectively.}
  \label{fig:square_signal}
\end{figure*}


\section{Example: Low-pass filter applied to square wave}\label{sec:micromagneticExample}
Before we conclude, we want to present an example of the application of the proposed skyrmion device.
As we have shown, the skyrmion device can filter out high frequencies from input signals, a key feature of RC circuits under AC currents. For this reason, it has the ability of the low-pass filter to transform waveforms. Therefore, we use the skyrmion device to transform a square wave into a square or triangular response, depending on the current frequency.

The results of micromagnetic simulations are presented in Fig.~\ref{fig:square_signal}(a), where we plot the skyrmion's position under an applied square wave current [or input $\xin(t)$, as shown in Fig.~\ref{fig:square_signal}(b)] versus the time for various frequencies $\omega = n \ws$. We vary $n$ and simulate up to $t= 5 T_{\mathrm{s}}/n$. Note that we have normalized the amplitude in Fig.~\ref{fig:square_signal}(a).

As shown in Figure~\ref{fig:square_signal}(a), the skyrmion's trajectory approaches a square wave for small current frequencies. Upon switching on a positive or negative current, the skyrmion moves quickly from one point to another and stays there for most of the time. This is because the time between switching the current is so large that the skyrmion can move to the respective saturation position where the resulting force from the changing PMA cancels the force induced by the STT. The results are similar to the charging and discharging characteristics presented for DC currents in Sec.~\ref{sec:micromagneticDC}. However, as the frequency increases, i.e., as the time between current switches shortens, the skyrmion's position gradually takes the form of a triangular wave.

It is worth noting that this triangular response can also be achieved with skyrmions in racetracks without spatial variations in PMA. In such a system, at all frequencies, the skyrmion moves uniformly in both directions. In our case, the skyrmion only moves slightly away from the lowest anisotropy for high frequencies. As a result, it is minimally influenced by the PMA variation and behaves as if it were in a homogenous racetrack.

The low-pass filter's characteristic feature is that the square wave can be transformed into a square or triangular response, depending on the current's frequency. This result has been achieved in Fig.~\ref{fig:square_signal}(a), but the transformation can be better quantified by expressing the signals as Fourier series
\begin{equation}
    x(t) = \sum_{k} b_k \sin\left(k \omega t\right),
\end{equation}
where $\omega$ is the current frequency and $b_k$ are the Fourier coefficients. Superscripts $S$ and $T$ will be added to indicate the square and triangular waveforms, respectively. For a perfect square wave, $b^{S}_k \propto \frac{1}{k}$, while for a perfect triangular wave, $b^{T}_k \propto \frac{1}{k^2}$. In both cases, $k$ are odd integer numbers.

In Figure~\ref{fig:square_signal}(c-h), we compare these analytical Fourier transforms with numerically calculated Fourier transforms of $\xin(t)$ (orange color) and $x_\mathrm{out}(t)$ (blue pallet colors) based on the micromagnetic simulations presented in Fig.~\ref{fig:square_signal}(a,b). Again, they are characterized by $b_k$, where $k$ can now be any real number. The coefficients $b_k$ are normalized by $b_1$ to compare the results for the different frequencies. For comparison, in each panel, the orange and green dashed lines are plotted to display the decay rates $1/k$ and $1/k^2$, corresponding to a perfect square and triangular wave, respectively. 

The input signal $\xin(t)$ is always a perfect square wave, so $b_k$ exhibits peaks only at odd integer values for $k$ that decay as $\propto 1/k$, regardless of the frequency. The numerically calculated coefficients $b_k$ of the output signal $x_\mathrm{out}(t)$ also have peaks at the same odd-integer values for $k$, but the coefficients are also non-zero in between. Furthermore, the peaks at odd integer values for $k$ display different behaviors depending on the input current frequency $\omega=n\wc$: For small frequencies, the peaks in $b_k$ align more closely with the orange dashed line. However, as the frequency increases, the peaks in $b_k$ progressively shift closer to the green dashed line, reflecting a greater attenuation of higher harmonics. This reflects the transition from the square wave response for small frequencies to the triangular wave for high frequencies [From (c) to (h)].

Based on our previously established analytical results in Sec.~\ref{sec:analytical}, this transition can be understood by multiplying the normalized transfer function $\transferfunc_{\mathrm{s}}=\frac{1}{\sqrt{1 + k^2n^2}}$ of a periodic current with frequency $k\omega$ to the Fourier coefficients $b_k$ of the square wave input (peaks proportional to $\frac{1}{k}$) and considering the limit $n \gg 1$ for $\omega\gg\ws$. This results in odd-integer peaks of the Fourier coefficients $b_k$ proportional to $\frac{1}{k^2}$, agreeing with a triangular wave.

In summary, analyzing micromagnetic simulations under an applied square wave input with various frequencies demonstrates that the output signal $x_{\mathrm{out}}(t)$ undergoes the expected transition from a square to a triangular waveform as the frequency increases. These results exemplify our device concept's ability to modulate input waveforms and act as a low-pass filter.


\section{Conclusions}\label{sec:conclusion}


We have demonstrated a skyrmion device concept that allows us to emulate an RC circuit. We have shown analytically and via micromagnetic simulations that the skyrmion's position is analogous to the voltage at the capacitor. Under DC currents, the skyrmion position mimics the charging and discharging behavior of the capacitor. Under AC currents, it acts as a low-pass filter. In other words, the skyrmion's oscillation amplitude is attenuated at higher frequencies, allowing it to transform waveforms and filter signals.

To achieve this characteristic behavior, the energy landscape in the skyrmion system must depend quadratically on the position $x$. We have engineered this potential via a quadratic dependence of the anisotropy $K_\mathrm{u}(x)$ on the position $x$. In an actual device, this parameter can be locally modulated via focused ion beam irradiation or variations in the FM layer thickness ~\cite{juge2021helium,kern2022deterministic,ahrens2023skyrmions,yu2016room, gowtham2016thickness}. Although there are more reports in the scientific literature of successfully engineering the PMA, in theory, other parameters like the DMI coefficient could also be modulated \cite{wells2017effect,balk2017simultaneous}, resulting in a similar behavior.

Our results are significant from a fundamental point of view and for applications. Beyond directly resembling the electronic RC circuit, our findings further consolidate the application of skyrmions in neuromorphic computing: Artificial neurons are ideally described by the LIF (Leaky-Integrate-Fire) model that is equivalent to an RC circuit~\cite{gerstner2014neuronal}. In other words, the device concept presented in this paper also serves as a prototype for a skyrmion-based artificial neuron. The existing studies on skyrmion-based artificial neurons, such as Ref.~\cite{li2017magnetic}, consider a skyrmion moving in a linear-varying PMA racetrack. Therefore, this neuromorphic device mimics the behavior of biological neurons qualitatively but does not resemble the LIF model quantitatively. Our results motivate artificial skyrmion-based neurons that resemble the presented RC circuit dynamics so that the device concept can be used to understand and improve the potential of artificial neurons based on skyrmions.

The advances in nanotrack engineering could position skyrmion systems as candidates for devices that aim to reproduce alternative fundamental electronic components. In the future, other spatial dependencies of the energy landscape can be analyzed. For example, when the PMA parameter $K_\mathrm{u}$ depends on $x$ with a higher order, an RC circuit is emulated for which the resistance depends on the voltage. 

In any case, replicating electronic devices based on skyrmions is an effort worthwhile because, as we have shown, the time constant $\tau_\mathrm{s}=\frac{b D_0 \alpha}{\pi d_z r_0^2 K_2}$ has values of a few ns and the cutoff frequency $\ws=\tau_\mathrm{s}^{-1}$ of about hundred GHz. Therefore, our results suggest that skyrmions could be an alternative to RC circuits that face practical challenges or are unable to operate at this time scale.

\subsection*{Acknowledgements}
This work is supported by SFB TRR 227 of Deutsche Forschungsgemeinschaft (DFG) and SPEAR ITN.
This project has received funding from the European Union’s Horizon 2020 research and innovation program under the Marie Skłodowska-Curie grant agreement No 955671. 
This work was supported by the EIC Pathfinder OPEN grant 101129641 Orbital Engineering for Innovative Electronics.
Membership in the International Max Planck Research School for Science and Technology of Nano-Systems is gratefully acknowledged. I.A. performed the simulations. B.G. and I.A. wrote the manuscript with significant input from all authors. I.A. prepared the figures. All authors discussed the results. B.G. and I.M. planned and supervised the project.

\bibliography{references.bib} 
\bibliographystyle{naturemag}

\end{document}